\documentclass[twocolumn,superscriptaddress,showpacs,preprintnumbers,amsmath,amssymb]{revtex4}

\usepackage{booktabs}
\usepackage{graphicx,color}
\usepackage{dcolumn}
\usepackage{bm}
\usepackage{CJK}
\usepackage{mathrsfs}
\usepackage[pdfstartview=FitH,
            CJKbookmarks=true,
            bookmarksnumbered=true,
            bookmarksopen=true,
            colorlinks,
            pdfborder=001,
            linkcolor=blue,
            anchorcolor=blue,
            citecolor=blue
            ]{hyperref}
\usepackage{doi}

\begin{document}
\begin{CJK}{UTF8}{gbsn}

\title{Skyrme functional with tensor terms from \textit{ab initio} calculations \\ of neutron-proton drops}

\author{Shihang Shen (申时行)}
\affiliation{Dipartimento di Fisica, Universit\`a degli Studi di Milano, Via Celoria 16, I-20133 Milano, Italy}
\affiliation{INFN, Sezione di Milano, Via Celoria 16, I-20133 Milano, Italy}

\author{Gianluca Col\`o \footnote{Email: Gianluca.Colo@mi.infn.it}}
\affiliation{Dipartimento di Fisica, Universit\`a degli Studi di Milano, Via Celoria 16, I-20133 Milano, Italy}
\affiliation{INFN, Sezione di Milano, Via Celoria 16, I-20133 Milano, Italy}

\author{Xavier Roca-Maza}
\affiliation{Dipartimento di Fisica, Universit\`a degli Studi di Milano, Via Celoria 16, I-20133 Milano, Italy}
\affiliation{INFN, Sezione di Milano, Via Celoria 16, I-20133 Milano, Italy}

\date{\today}

\begin{abstract}
  A new Skyrme functional devised to account well for standard nuclear properties as well as for spin and spin-isospin properties is presented. The main novelty of this work relies on the introduction of tensor terms guided by \textit{ab initio} relativistic Brueckner-Hartree-Fock calculations of neutron-proton drops. The inclusion of tensor term does not decrease the accuracy in describing bulk properties of nuclei, experimental data of some selected spherical nuclei such as binding energies, charge radii, and spin-orbit splittings can be well fitted. The new functional is applied to the investigation of various collective excitations such as the Giant Monopole Resonance (GMR), the Isovector Giant Dipole Resonance (IVGDR), the Gamow-Teller Resonance (GTR), and the Spin-Dipole Resonance (SDR). The overall description with the new functional is satisfactory and the tensor terms are shown to be important particularly for the improvement of the Spin-Dipole Resonance results. Predictions for the neutron skin thickness based on the non-energy weighted sum rule of the Spin-Dipole Resonance are also given.\\
\end{abstract}


\maketitle

\section{Introduction}

Nuclear energy density functionals such as Skyrme, Gogny, and relativistic ones have achieved great success since a few decades \cite{Vautherin1972,Decharge1980,Bender2003,Meng2016}. Actually, by fitting to a reduced set of experimental data (mainly binding energies and charge radii), the obtained functionals can give very good description of ground-state as well as excited state properties (such as the excitation energies of Giant Resonances) along the whole nuclear chart. On the other hand, it should also be noted that there are still many open questions regarding the existing functionals. For example, certain properties are difficult to be tightly constrained from experimental data: one example is the symmetry energy \cite{Baldo2016,Roca-Maza2018,Roca-Maza2018a}, that is at the same time extremely relevant to describe exotic neutron-rich nuclei or neutron stars. 
Another key point is that there is not a unique ansatz to write down the structure of a functional; neither one knows clearly which is the path for the systematic improvement of existing functionals. In the context of the present paper, it is relevant to stress that the role of tensor terms within nuclear DFT is not completely elucidated despite significant efforts and partial achievements 
\cite{Sagawa2014}.

It is well known that the tensor force is an important component of the bare nucleon-nucleon interaction \cite{Brown2010}, but its relevance for energy density functionals is another question. 
In earlier implementations of DFT, based on effective Hamiltonians, the tensor force was either deemed not important or often ignored for the sake of simplicity.
Nowadays, the situation has changed in particular for what concerns the Skyrme and Gogny functionals 
(the discussion about relativistic functionals is outside our scope here but the reader can consult 
Refs. \cite{Sagawa2014,Meng2016}). 
Skyrme functionals have the special property that the contribution of the tensor force to the functional
goes together with that from the exchange terms of the central force, if we start from an
effective interaction (or pseudo-potential). These contributions are called $\mathbf{J}^2$ terms of
the functional (see Sec.~\ref{sec:theory}, where we define precisely the so-called spin-orbit
density $\mathbf{J}$). Since these terms vanish in spin-orbit saturated nuclei, and many nuclei
used in the fits belong to this category, one can easily understand  
why not only the tensor terms but the whole $\mathbf{J}^2$ terms were ignored in most of the 
early Skyrme parametrizations (see, for example, the discussion in Ref.~\cite{Lesinski2007}).

The revival of the discussion on tensor terms occurred in the beginning of the 21st century. With the development of new radioactive ion beam facilities, more and more exotic neutron-rich nuclei have been studied and new phenomena regarding the evolution of the spin-orbit splittings with neutron excess have been highlighted \cite{Schiffer2004}. Soon afterwards, Otsuka \textit{et al.} suggested the importance of an effective neutron-proton tensor force to explain this evolution 
\cite{Otsuka2005}. A number of experimental data could be reproduced, to a certain extent better by including tensor terms, within the Gogny \cite{Otsuka2006}, Skyrme \cite{Brown2006,Colo2007,Brink2007,Lesinski2007}, and relativistic frameworks \cite{LongWH2008} (cf. also Refs. \cite{Jiang2015a,LiJJ2016,Karakatsanis2017,Brink2018,Lopez-Quelle2018,ZongYY2018,WangZH2018}).

In general, however, it is difficult to single out observables that can uniquely pinpoint the effects of the tensor force. For example, if one looks at single-particle data, it is well known that particle-vibration coupling (PVC) is also important for single-particle properties \cite{Colo1994,NiuYF2012,Litvinova2011,Afanasjev2015a}; as a consequence, trying to determine the strength of the tensor terms by looking at experimental single-particle properties without including PVC effects in the theory, may not be well justified. 
This may explain the difficulties found in the systematic study carried out in Ref.~\cite{Lesinski2007}, where 
36 Skyrme functionals were produced, characterized by different strengths of tensor terms, and properties like single-particle levels, spin-orbit splittings, mass residuals and anomalies of radii were compared with experimental data. 
It has been found very hard to satisfy all the experimental constraints, so that the Skyrme ansatz itself 
has been blamed for this deficiency.

In such a situation, \textit{ab initio} calculations should become a benchmark. Based on the use of realistic interactions which can describe two-nucleon scattering phase shifts and deuteron properties, \textit{ab initio} calculations have no free parameters and can provide in some cases a reliable connection between few-body and many-body physics \cite{Drut2010}. The obtained results can then be used as pseudodata, in order to complement the experimental data and calibrate nuclear density functionals. For example, pseudodata obtained within the Brueckner-Hartree-Fock (BHF) theory \cite{Day1967} do not contain effects like particle-vibration coupling, and can be used directly as a benchmark for DFT calculations.

In the case of electronic systems, the derivation of energy density functionals from \textit{ab initio} calculations has been developed for a long time and has achieved many successes \cite{Perdew2003}. Because of the more complicated nature of the nuclear interaction as compared with the Coulomb interaction, \textit{ab initio} calculations in nuclear physics are much more difficult \cite{Dickhoff2004,Lee2009,LiuL2012,Barrett2013,Hagen2014,Carlson2015,Hergert2016,Shen2016,Shen2017} and fall behind that in atomic physics. 
In fact, the study of the nucleon-nucleon interaction itself, especially its derivation from the 
QCD-inspired chiral Effective Field Theory (EFT) vs. the traditional meson-exchange force, or the 
precise assessment of three-body forces, is a very active 
and important field of research where the community is still making progress now \cite{Epelbaum2009,Machleidt2011}. Thus, despite the difficulties, one can say that the nuclear \textit{ab initio} calculations can provide valuable information for the derivation of nuclear energy density functionals 
\cite{Drut2010,LiangHZ2018}. 

In this context, 
recently, studies on the neutron drops are receiving more and more attention \cite{Pudliner1996,Gandolfi2011,Maris2013,Potter2014,Tews2016,ZhaoPW2016,Shen2018,Shen2018b,Bonnard2018}.
The neutron drop is an ideal system composed of a finite number of neutrons confined in an external field.
Similarly as nuclear matter, which is also an ideal and relatively easy system due to translational 
invariance that yet can provide us with valuable information \cite{Roca-Maza2018a}, neutron drops are 
also simple (only neutron-neutron interaction, no center-of-mass correction because of the external field) 
and can give an interesting insight on shell structure and/or finite-size properties, which is not the case 
for nuclear matter.
Neutron systems are also more favorable for \textit{ab initio} treatment, due also to the large
value of the neutron-neutron scattering length that makes the system close to the unitary limit \cite{braaten2006}.

For example, by comparing the \textit{ab initio} quantum Monte-Carlo (QMC) calculations with Skyrme Hartree-Fock calculations, suggestions for the improvement of future Skyrme energy density functionals 
have been given concerning the density distribution, the spin-orbit splittings, the 
isovector gradient terms of the functionals, and the pairing sector \cite{Pudliner1996,Gandolfi2011}.
In another work, the properties of periodically modulated neutron matter were studied by the quantum Monte-Carlo method, and this was used to constrain the isovector term of Skyrme functionals \cite{Buraczynski2016,Buraczynski2017}.
In Ref.~\cite{Bonnard2018}, the \textit{ab initio} calculations for neutron drops are used to further constrain previous EFT-inspired energy-density functionals in nuclear matter, by comparing quantities 
like the total energies, single-particle potentials, and density distributions.
In a recent work \cite{Shen2018,Shen2018b}, the effect of the tensor force in neutron drops has 
been studied by the \textit{ab initio} relativistic Brueckner-Hartree-Fock (RBHF) theory with the 
Bonn A interaction \cite{Machleidt1989}.
A clear evidence of the impact of tensor force on the evolution of spin-orbit splittings has been
found, and this provides a guide for determining the strength of the tensor force in the nuclear medium \cite{Shen2018,Shen2018b}.

In these works~\cite{Shen2018,Shen2018b}, only information on the tensor force between neutrons is present, but not information on the interaction between neutrons and protons. This information is essential for a complete study of the tensor force in nuclei, where all the previous studies we have mentioned point to the dominance of the neutron-proton component. Therefore, in this work we include the protons into the neutron drops, that is, we study the neutron-proton drops with the RBHF theory. Obviously, we are not comparing with any experimental data, but conceptually we deem we can use the results on this ideal system as pseudo-data to provide reliable information 
to constrain a nuclear energy density functional. In this work we aim at fitting a Skyrme energy density functional. We keep this neutron-proton system as simple as possible: we do not include the Coulomb interaction between the protons, and neutron and proton masses are the same.

There are two main reasons why we choose to study artificial trapped neutron-proton drops instead of realistic finite nuclei. First, concerning the purpose of this work, there is no difference between extracting information on the tensor force from finite nuclei and from artificial neutron-proton drops. But, without center-of-mass correction and Coulomb interaction, the neutron-proton drops are easier to be calculated within RBHF. Secondly, following the spirit of using ideal systems such as nuclear matter and neutron-drops to study different aspects of nuclear systems, the neutron-proton drops provide another ideal case to investigate different properties with more freedom than finite nuclei: for instance, one can vary the neutron and proton numbers without worrying if the system is bound or not.

At variance with previous works in which the tensor force is added perturbatively on top of 
existing Skyrme functionals, such as in Ref.~\cite{Colo2007}, the tensor force in this work is included self-consistently with a refit of the whole parameter set, like in the case of the TIJ family~\cite{Lesinski2007}. But differently from the case of the TIJ family, 
the tensor terms in this work will be fitted using the pseudo-data of the neutron-proton drops coming from RBHF calculations.
Of course, such pseudo-data does not necessarily need to come from RBHF calculations or from a specific \emph{ab initio} calculation, relativistic or non-relativistic. In principle, relativistic effects such as nucleon-antinucleon excitations can be expressed in nonrelativistic framework in terms of a three-body force \cite{Brown1987}, and it has been shown that the results of RBHF using the one-boson-exchange Bonn interaction does agree well with the results of nonrelativistic BHF using two-body and three-body forces \cite{Sammarruca2012}.
For the fit as a whole, we take inspiration from the successful fitting protocol of SAMi \cite{Roca-Maza2012,Roca-Maza2013b}, which has been shown to display satisfactory results for spin and spin-isospin resonances. The newly established Skyrme functional is named SAMi-T. 

In Sec. \ref{sec:theory}, we give a brief summary of the Skyrme interaction with tensor terms, and Hartree-Fock equations. The numerical details for the fitting are discussed in Sec. \ref{sec:nd}. Results with the new interaction for nuclear matter and finite nuclei are presented in Sec. \ref{sec:res}. Finally, the summary and perspectives for future investigations will be given in Sec. \ref{sec:sum}.


\section{Theoretical Framework}\label{sec:theory}

\subsection{Relativistic Brueckner-Hartree-Fock theory}

In this Subsection, we will outline the theoretical framework of the RBHF for finite nuclear system.
For more detail one can see Refs.~\cite{Shen2016,Shen2017}.

We start with a relativistic one-boson-exchange $NN$ interaction which describes the $NN$ scattering data~\cite{Machleidt1989}. The Hamiltonian can be expressed as:
\begin{equation}
H=\sum_{kk^{\prime }}\langle k|T|k^{\prime }\rangle a_{k}^{\dagger
}a_{k^{\prime }}^{{}}+\frac{1}{2}\sum_{klk^{\prime }l^{\prime }}\langle
kl|V|k^{\prime }l^{\prime }\rangle a_{k}^{\dagger }a_{l}^{\dagger
}a_{l^{\prime }}^{{}}a_{k^{\prime }}^{{}},  \label{eq:hami}
\end{equation}%
where the relativistic matrix elements are given by
\begin{align}
\langle k|T|k^{\prime }\rangle & =\int d^{3}r\,\bar{\psi}_{k}(\mathbf{r})\left( -i%
\bm{\gamma}\cdot \nabla +M\right) \psi _{k^{\prime }}(\mathbf{r}), \\
\langle kl|V|k^{\prime }l^{\prime }\rangle & = \sum_{\alpha} \int
d^{3}r_{1}d^{3}r_{2}\,\bar{\psi}_{k}(\mathbf{r}_{1})\Gamma _{\alpha }^{(1)}\psi
_{k^{\prime }}(\mathbf{r}_{1}) \label{eq:V} \notag \\
&~~~~~~~~\times D_{\alpha }(\mathbf{r}_{1},\mathbf{r}_{2})\bar{\psi}_{l}(\mathbf{r}_{2})\Gamma
_{\alpha }^{(2)}\psi _{l^{\prime }}(\mathbf{r}_{2}).
\end{align}
The indices $k,l$ run over a complete basis of Dirac spinors with positive and negative energies.
The two-body interaction $V$ contains the contributions from different mesons labelled by $\alpha$: scalar $(\sigma, \delta)$, vector $(\omega, \rho)$, and pseudovector $(\eta,\pi)$. For the isovector mesons $\delta, \rho,$ and $\pi$ additional isospin matrices $\vec\tau$ are included.
In the Bonn interaction \cite{Machleidt1989}, a form factor of monopole-type is attached to each vertex and $D_{\alpha }(\mathbf{r}_{1},\mathbf{r}_{2})$ is the meson propagator.
Retardation effects were deemed to be small and omitted from the beginning.

In Brueckner's theory \cite{Brueckner1954_PR95-217,Day1967}, the effective interaction in the nuclear medium, the $G$-matrix, is used instead of the bare nucleon-nucleon interaction, which has a strong repulsive core and is difficult to be used directly in nuclear many-body theory.
The $G$-matrix takes into account the short-range correlations by summing up all the ladder diagrams of the bare interaction and it is obtained by solving the Bethe-Goldstone equation,
\begin{equation}
\bar{G}_{aba'b'}(W) = \bar{V}_{aba'b'}
 +\frac{1}{2}\sum_{cd} \frac{\bar{V}_{abcd}\bar{G}_{cda'b'}(W)}{W-e_{c}-e_{d}},
\label{eq:BG}
\end{equation}
where in the RBHF theory $|a\rangle,|b\rangle$ are eigenstates of the relativistic Hartree-Fock equations with $e_{a}, e_{b}$ the corresponding single-particle energies, $\bar{V}_{aba'b'}$ are the anti-symmetrized two-body matrix elements (\ref{eq:V}).
The intermediate states $c,\,d$ run over all states above the Fermi surface with $e_c,\,e_d > e_F$.
In the above expression, $W$ is the starting energy \cite{Baranger1969_Varenna40,Davies1969_PRC177-1519,Shen2017}.

The single-particle motion fulfills the relativistic Hartree-Fock equation in the external field of a harmonic oscillator (HO):
\begin{equation}
(T+U+U_{\rm HO})|a\rangle =e_{a}|a\rangle,
\label{eq:rhf}
\end{equation}%
with HO potential
\begin{equation}\label{eq:UHO}
  U_{\rm HO}(\mathbf{r}) = \frac{1}{2}m\omega^2 r^2,
\end{equation}
where $m$ is the nucleon mass and $\hbar\omega$ is the HO strength.
The self-consistent single-particle potential $U$ is calculated by the $G$-matrix
\begin{equation}
\langle a|U|b\rangle = \sum_{c=1}^{A}\langle ac|\bar{G}|bc\rangle ,
\label{eq:UG}
\end{equation}
where the index $c$ runs over the occupied states in the Fermi sea (\emph{no-sea} approximation).

In the end, the total energy is calculated as
\begin{equation}
  E_{\rm RBHF} = \sum_{a=1}^A \langle a|T|a\rangle + \frac{1}{2} \sum_{a,b=1}^A \langle ab|\bar{G}|ab\rangle + \langle U_{\rm HO}\rangle.
\end{equation}
where $\langle U_{\rm HO}\rangle$ is the expectation value of the external field.

\subsection{Skyrme density functional}

In this Subsection, we will outline the theoretical framework of the Skyrme interaction and the Skyrme-Hartree-Fock theory.
For a detailed description of the Skyrme Hartree-Fock theory and the formulas in spherical nuclei, we refer to~\cite{Vautherin1972}.

The Skyrme effective interaction with two-body tensor force is written in the standard form as~\cite{Vautherin1972,Stancu1977}.
\begin{widetext}
\begin{align}
V(\mathbf{r}_1,\mathbf{r}_2) &= t_0(1+x_0P_\sigma) \delta(\mathbf{r}) + \frac{1}{2}t_1
(1+x_1P_\sigma) \left[ {\mathbf{P}'}^2\delta(\mathbf{r}) + \delta(\mathbf{r}) \mathbf{P}^2 \right]
+ t_2(1+x_2P_\sigma) \mathbf{P}' \cdot \delta(\mathbf{r}) \mathbf{P},  \notag \\
&~~~+ \frac{1}{6}t_3 (1+x_3P_\sigma) \rho^\gamma(\mathbf{R}) \delta(\mathbf{r})
+ iW_0(\bm{\sigma}_1+\bm{\sigma}_2) \cdot \left[ \mathbf{P}'\times \delta(\mathbf{r}) \mathbf{P} \right] + V_T(\mathbf{r}_1,\mathbf{r}_2), \label{eq:vskyrme} \\
V_T(\mathbf{r}_1,\mathbf{r}_2) &= \frac{T}{2} \left\{\left[(\bm{\sigma}_1\cdot\mathbf{P}')
(\bm{\sigma}_2\cdot\mathbf{P}')-\frac{1}{3}(\bm{\sigma}_1\cdot\bm{\sigma}_2)
{\mathbf{P}'}^{2}\right]\delta(\mathbf{r})
+\delta(\mathbf{r})\left[(\bm{\sigma}_1\cdot\mathbf{P})
(\bm{\sigma}_2\cdot\mathbf{P})-\frac{1}{3}(\bm{\sigma}_1\cdot\bm{\sigma}_2)
{\mathbf{P}}^{2}\right]\right\} \notag \\
&~~~+U\left\{(\bm{\sigma}_1\cdot\mathbf{P}')\delta(\mathbf{r})
(\bm{\sigma}_2\cdot\mathbf{P})-\frac{1}{3}(\bm{\sigma}_1\cdot\bm{\sigma}_2)
\left[\mathbf{P}'\cdot\delta(\mathbf{r})\mathbf{P}\right]\right\}, \label{eq:vt}
\end{align}
\end{widetext}
where $\mathbf{r} = \mathbf{r}_1 - \mathbf{r}_2, \mathbf{R} = \frac{1}{2}(\mathbf{r}_1 + \mathbf{r}_2), \mathbf{P} = \frac{1}{2i}(\nabla_1-\nabla_2)$, $\mathbf{P}'$ is the hermitian conjugate of $\mathbf{P}$ acting on the left.
The spin-exchange operator reads $P_\sigma = \frac{1}{2}(1+\bm{\sigma}_1\cdot\bm{\sigma}_2)$, and $\rho$ is the total nucleon density.

The Hartree-Fock equations can be obtained by the variational method once the total energy of 
the Hartree-Fock ground state is written down. The Hartree-Fock equations read
\begin{equation}\label{eq:}
  \left[ -\frac{\hbar^2}{2m}\nabla^2 + U_q(\mathbf{r}) \right] \psi_k(\mathbf{r}) = e_k \psi_k(\mathbf{r}),
\end{equation}
where $e_k$ is the single-particle energy, $\psi_k$ is the corresponding wave function, 
and $q = 0(1)$ labels neutrons (protons).
The single-particle potential $U_q(\mathbf{r})$ is a sum of central, Coulomb and spin-orbit terms,
\begin{equation}\label{eq:}
  U_q(\mathbf{r}) = U_{q}^{\rm(c)}(\mathbf{r}) + \delta_{q,1}U_{C}(\mathbf{r}) + \mathbf{U}_{q}^{\rm(s.o.)}(\mathbf{r}) \cdot (-i)(\nabla\times\bf{\sigma}).
\end{equation}
The detailed expression of the central and Coulomb terms can be found, e.g., in Ref.~\cite{Chabanat1998}.
For the calculation of neutron(-proton) drops, we add an external field, which in this work is chosen as the harmonic oscillator potential (\ref{eq:UHO}).

The spin-orbit term reads \cite{Stancu1977,Reinhard1995,Sagawa2014}
\begin{equation}\label{eq:Uso}
  \mathbf{U}_q^{\rm(s.o.)}(\mathbf{r}) = \frac{1}{2} \left[ W_0\nabla\rho + W_0'\nabla\rho_q \right]
  + \left[ \alpha \mathbf{J}_q + \beta\mathbf{J}_{1-q} \right],
\end{equation}
where $\mathbf{J}(\mathbf{r})$ the spin-orbit density~\cite{Chabanat1998}. 
One should note that, starting from Eq.~(\ref{eq:vskyrme}) one would derive $W_0' = W_0$.
We adopt here a more general form and release this constraint as in SAMi \cite{Roca-Maza2012}.
The parameters $\alpha$ and $\beta$ in Eq.~(\ref{eq:Uso}) include contributions from 
the (exchange part of the) central force and from the tensor force,
\begin{equation}\label{eq:alpha-beta}
  \alpha = \alpha_c + \alpha_T,\quad \beta = \beta_c + \beta_T.
\end{equation}
They are given by
\begin{subequations}\label{eq:}\begin{align}
  \alpha_c &= \frac{1}{8}(t_1-t_2) - \frac{1}{8}(t_1x_1+t_2x_2),\quad
  \alpha_T = \frac{5}{12}U,\\
  \beta_c &= -\frac{1}{8}(t_1x_1+t_2x_2),\quad
  \beta_T = \frac{5}{24}(T+U).
\end{align}\end{subequations}

In the end, the Hartree-Fock total energy can be calculated in the usual way 
and reads
\begin{equation}\label{eq:}
  E_{\rm HF} = \frac{1}{2}\sum_i^A (t_i+e_i) + E_{\rm rear} + \frac{1}{2} \langle U_{\rm HO}\rangle,
\end{equation}
where $t_i$ is the expectation value of the single-particle kinetic energy on the orbit $i$, 
$E_{\rm rear}$ is the rearrangement term.
The factor one half in front of the external field arises because the remaining contribution is included in $e_i$, in full analogy with the kinetic energy which is also a one-body operator.

\section{Numerical details}\label{sec:nd}

In the new Skyrme functional established in this work, we include the so called $\mathbf{J}^2$ term, including $\alpha_c$ and $\beta_c$ from the central force, $\alpha_T$ and $\beta_T$ from the tensor force, as shown in Eq.~(\ref{eq:alpha-beta}). The Hartree-Fock equation is solved in coordinate space with a spherical box having size $R = 15$ fm and a radial mesh of $0.1$ fm. The fitting protocol is based on the successful SAMi functional \cite{Roca-Maza2012}, with further constraint on the tensor force from neutron-proton drops by RBHF theory: more precisely, the pseudo-data of neutron matter from \textit{ab initio} calculations are replaced by the results of neutron drops from RBHF calculations. The set of data and pseudodata to be fitted are:
\begin{enumerate}
  \item the binding energies $B$ and charge radii $r_c$ of $^{40}$Ca, $^{48}$Ca, $^{90}$Zr, $^{132}$Sn, and $^{208}$Pb;
  \item the spin-orbit splittings $\Delta E^{\rm s.o.}$ of the proton $1d$ orbit in $^{40}$Ca, proton $1g$ orbit in $^{90}$Zr, and proton $2f$ orbit in $^{208}$Pb.
  This will mainly determine the spin-orbit term $W_0$ and $W_0'$;
  \item the relative change of spin-orbit splittings from neutron-proton drops $^{40}20 (Z=20,N=20)$ to $^{48}20$ calculated by RBHF theory using Bonn A interaction.
  Since only the relative change is fitted, it mainly determines the tensor term $\alpha_T$ and $\beta_T$ but not the spin-orbit term $W_0$ and $W_0'$;
  \item the total energy of neutron drops with neutron number $N = 8, 20, 40, 50$ in a $\hbar\omega = 10$ MeV harmonic oscillator (HO) field calculated by RBHF using Bonn A interaction \cite{Shen2018,Shen2018b};
  \item we also keep the empirical hierarchy of the spin $G_0$ and spin-isospin $G_0^\prime$ Landau-Migdal parameters: $G_0^\prime > G_0>0$ \cite{Suzuki1999,Wakasa2005}.
\end{enumerate}

As mentioned in Ref.~\cite{Shen2018}, the evolution of the spin-orbit splittings by RBHF theory can be used to constrain the tensor force. On the other hand, the spin-orbit splittings given by RBHF theory using Bonn A interaction are slightly smaller than experimental data \cite{Shen2016,Shen2017}. For example, proton $1p$ splitting of $^{16}$O given by RBHF is $5.4$ MeV, in comparison with experimental value, that is $6.2$ MeV. Therefore in our fit, we use the relative change of spin-orbit splittings of neutron-proton drops to constrain the tensor force, and the experimental data of finite nuclei to constrain the absolute strength of spin-orbit splittings.

For the constraint on the isovector channel, we use the total energy of neutron drops calculated by RBHF instead of the total energy of neutron matter 
as in the case of SAMi. The results of RBHF are in reasonable agreement with other \textit{ab initio} calculations and show some advantages: for $N\leq 14$ the results of RBHF are similar to the results of AV8' + IL7, but for $N > 14$ RBHF gives more repulsion comparing with AV8' + IL7 \cite{Shen2018}. This is favourable as AV8' + IL7 can describe well the light nuclei system but tends to give too much attraction for pure neutron systems \cite{Sarsa2003} (see also green shadowed area and red squares in Fig. \ref{fig:ea-nd} and related text below). This is also more consistent for our purpose as the tensor force is constrained from the same RBHF calculations.

The numerical details of RBHF calculations can be found in Refs.~\cite{Shen2017,Shen2018a,Shen2018b} and briefly summarized here. The calculations are performed in spherical box with box size $R = 7$ fm.
Notice the box size for RBHF calculations is considerably smaller than that of Skyrme functional calculations, which is 15 fm. In principle, a very large box would be the best choice, but in practice one needs to find a balance between the precision and computational cost. For RBHF calculations, the convergence properties with respect to the box size have been carefully checked in Ref.~\cite{Shen2017} for medium-heavy nuclei, and in going from box size $R = 7$ fm to $R = 8$ fm, the energy per nucleon has been found to change less than $0.07$ MeV. With an additional external HO field, this convergence will be further improved as the single-particle wave functions decay very fast when $r$ goes out of the size of the HO potential. Therefore, for our purpose here, dealing with $50$ nucleons at most in a HO trap (the rms radii for a $N = 50$ neutron drop is $3.61$ fm), the box size $R = 7$ fm is enough. As for Skyrme functional calculations, since heavy nuclei such as $^{208}$Pb are also to be studied, a larger box size is needed.

The initial single-particle basis is chosen as a Dirac Woods-Saxon basis \cite{Zhou2003}, and the energy cut-off is $1100$ MeV for the positive energy states and $-1700$ MeV for the negative energy states (with a supplementary condition that at least 2 negative-energy states are included in each $(l,j)$ block). The single-particle angular momentum cut-off is 24 $\hbar$. For neutron-proton drops with $N = 20$, the two-particle coupled total angular momentum cut-off is 8 $\hbar$; for $N = 28$, this is 10 $\hbar$. If not specified otherwise, in all the calculations of this work the external field is chosen as an harmonic oscillator field with $\hbar\omega = 10$ MeV. For RBHF, it has been implemented as a vector external field.

The results of neutron-proton drops calculated by RBHF theory are shown in Table~\ref{tab:so-npd}, including the total energies, neutron rms radii, and spin-orbit splittings. From the Table one can see the effect of tensor force on the evolution of spin-orbit splittings as the number of neutron changes \cite{Otsuka2005,Shen2018}. For the system with $N = 20$, that is spin saturated ($1d$ orbit fully occupied), the tensor force makes no effect; for the system with $N = 28$, that is spin unsaturated (the $1f_{7/2}$ orbit is occupied while the $1f_{5/2}$ is empty), the tensor force produces the largest effect so that the spin-orbit splittings decrease.

\begin{table}[!th]
\caption{Total energies, neutron rms radii, neutron $(\nu)$ and proton $(\pi)$ $1p$ and $1d$ spin-orbit splittings of neutron-proton drops calculated by RBHF theory with Bonn A interaction.
All energies are in MeV and radii in fm.}
\label{tab:so-npd}
\centering
\begin{ruledtabular}
\begin{tabular}{rrrrrrr}
& & & \multicolumn{2}{c}{$\Delta E_{\nu}^{\rm s.o.}$} & \multicolumn{2}{c}{$\Delta E_{\pi}^{\rm s.o.}$}  \\
$(N,Z)$   & $E$    & $r_n$ & $1p$ & $1d$ & $1p$ & $1d$ \\
\hline                                                        
$(20,20)$ &  68.6 & 2.82  & 5.52 & 8.73 & 5.52 & 8.73 \\
$(28,20)$ & 162.1 & 3.02  & 2.91 & 5.82 & 2.85 & 4.93 \\
\end{tabular}
\end{ruledtabular}
\end{table}

\section{Results and discussion}\label{sec:res}

\subsection{Fitted results}

In Table~\ref{tab:datafit} we list the data and pseudodata $\mathcal{O}_i$ to be fitted, together with the corresponding errors for the fit $\Delta \mathcal{O}_i$ and the number of data points $n_{\rm data}$. The partial contributions to the total $\chi^2$ from each category are also listed.
The error for total binding energy is chosen instead of binding energy per nucleon. This will give more weight to heavier nuclei.
Again, as in the fitting protocol of SAMi \cite{Roca-Maza2012}, the errors of the pseudodata are chosen in such a way that these data are used as a guide for the new functional and shall not affect the accuracy of fitting to experimental data. From the Table, it can be seen that when including tensor, the description of the experimental binding energies and charge radii keeps similar accuracy as for other Skyrme functionals.

\begin{table}[!ht]
  \caption{Data and pseudodata $\mathcal{O}_i$, adopted errors for the fit $\Delta \mathcal{O}_i$, partial contributions to the total $\chi^2$, and number of data points $n_{\rm data}$.}
  \label{tab:datafit}
  \centering
  \begin{ruledtabular}
  \begin{tabular}{lcrrr}
  $\mathcal{O}_i$ & $\Delta \mathcal{O}_i$ & $\chi_{\rm partial}^2$ & $n_{\rm data}$ & Ref. \\
  \hline
  $B$   & $1.00$ MeV & $ 9.71$ & 5 & \cite{WangM2017} \\
  $r_c$ & $0.01$ fm  & $20.09$ & 5 & \cite{Angeli2013} \\
  $\Delta E_{\rm s.o.}$ & $0.04 \times \mathcal{O}_i$ & $11.40$ & 3 & \cite{Zalewski2008} \footnote{data compiled in table III of Ref.~\cite{Zalewski2008}, see caption of this table for further details.} \\
  $\Delta(\Delta E_{\rm s.o.}^{\rm RBHF})$ & $0.01 \times \mathcal{O}_i$ & $3.25$ & 4 & [Tab. \ref{tab:so-npd}]\\
  $E_N^{\rm RBHF}/(\hbar\omega N^{4/3})$ & $0.01$ & $3.02$ & 4 & \cite{Shen2018} \\
  \hline
  $\chi^2$ & & \multicolumn{3}{c}{$47.5/21 = 2.26$} \\
  \end{tabular}
  \end{ruledtabular}
\end{table}

The parameters and saturation properties of SAMi-T functional are listed in Table~\ref{tab:para}, where the corresponding estimations of the standard deviation \cite{Bevington1992} are also given. For the detail of the covariance analysis, see Appendix \ref{app:err}.
Comparing with SAMi \cite{Roca-Maza2012}, the strengths of the spin-orbit terms for SAMi-T are larger (for SAMi, $W_0 = 137$ MeV fm$^{5}$, $W_0' = 42$ MeV fm$^{5}$). This is understandable as SAMi does not contain a genuine tensor force that is known to reduce the spin-orbit splittings of spin-unsaturated systems such as $^{90}$Zr or $^{208}$Pb that have been fitted in both functionals. Therefore, the new set including tensor needs larger spin-orbit 
strength to reproduce the same data.

\begin{table}[!ht]
  \caption{SAMi-T parameter set and saturation properties with estimated standard deviations.}
  \label{tab:para}
  \centering
  \begin{ruledtabular}
  \begin{tabular}{lcclc}
  & Value & Error & & Value \\
  \hline
  $t_0$      & $ -2199.38$ MeV fm$^{3}$         & $ 372.$ & $\rho_0$ & $0.164(1)$ fm$^{-3}$ \\
  $t_1$      & $  533.036$ MeV fm$^{5}$         & $ 20.7$ & $e_0$    & $-16.15(3)$ MeV \\
  $t_2$      & $ -88.1692$ MeV fm$^{5}$         & $ 12.6$ & $m_{\rm IS}^*/m$ & $0.634(19)$ \\
  $t_3$      & $  11293.5$ MeV fm$^{3+3\gamma}$ & $2014.$ & $m_{\rm IV}^*/m$ & $0.625(122)$ \\
  $x_0$      & $ 0.514710$                      & $0.178$ & $J$ & $29.7(6)$ MeV \\
  $x_1$      & $-0.531674$                      & $0.593$ & $L$ & $46(12)$ MeV \\
  $x_2$      & $-0.026340$                      & $0.117$ & $K_0$ & $244(5)$ MeV \\
  $x_3$      & $ 0.944603$                      & $0.481$ & $G_0$ & $0.08$ (fixed) \\
  $\gamma$   & $ 0.179550$                      & $0.047$ & $G_0'$ & $0.29$  (fixed)\\
  $W_0$      & $  130.026$ MeV fm$^5$           & $  8.2$ \\
  $W_0'$     & $  101.893$ MeV fm$^5$           & $ 18.6$ \\
  $\alpha_T$ & $ -39.8048$ MeV fm$^5$           & $ 39.9$ \\
  $\beta_T$  & $  66.6505$ MeV fm$^5$           & $ 39.9$ \\
  \end{tabular}
  \end{ruledtabular}
\end{table}

It is worth to discuss in a specific manner 
the values of the so-called Landau-Migdal parameters $G_0$ and $G_0'$ which are associated, respectively, with the spin and spin-isospin particle-hole (p-h) interaction. If one only looks at the ground state properties, different sets with different values of $G_0$ and $G_0'$ can achieve similar accuracy \cite{Cao2010}. 
But if one wants to have at the same time a good description of the properties of spin and spin-isospin excitations like the Gamow-Teller or spin-dipole resonance, there will be more restrictions \cite{Roca-Maza2012}. As suggested in Refs. \cite{Suzuki1999,Wakasa2005}, and taken into account by the fit of SAMi, we ought to respect the relations $G_0' > G_0 > 0$. Here, we obtain slightly different values with respect to SAMi as we are attempting to include more terms in the functional (tensor terms) and to reconcile with different pseudo-data (associated with neutron-proton drops and neutron drops from RBHF calculations).

\subsection{Ground state}

In Table~\ref{tab:erso-fn} we give the binding energy $B$ and charge radius $r_c$ of several doubly magic spherical nuclei calculated by using SAMi-T, in comparison with experimental data \cite{WangM2017,Angeli2013} when existing. In most cases the descriptions of binding energy and charge radius are accurate within $1\%$.
Considering that there are $11$ free parameters ($G_0$ and $G_0'$ have been fixed), this accuracy is comparable with other commonly seen functionals, nonrelativistic or relativistic ones. For example, as indicated by the fitted $\chi^2$ , the $\chi^2$/data of SAMi-T for binding energies and radii are $1.9$ and $4.0$, respectively (Table \ref{tab:datafit}), while those of SAMi are $6.5$ and $3.3$ with similar data sets \cite{Roca-Maza2012}.

\begin{table}[!th]
  \caption{Binding energy $B$ and charge radius $r_c$ of several doubly magic spherical nuclei calculated by using SAMi-T (with estimated standard deviations in the parentheses), in comparison with experimental data \cite{WangM2017,Angeli2013}.}
  \label{tab:erso-fn}
  \centering
  \begin{ruledtabular}
  \begin{tabular}{lrrrrr}
  El. & $A$ & $B$ (MeV) & $B^{\rm expt}$ (MeV) & $r_c$ (fm) & $r_c^{\rm expt}$ (fm) \\
  \hline
  O  &  16 &  127.78(33)   &  127.62 & 2.774(4) & 2.699 \\
  Ca &  40 &  343.74(52)   &  342.05 & 3.477(3) & 3.478 \\
     &  48 &  415.32(50)   &  415.99 & 3.515(3) & 3.477 \\
  Ni &  56 &  468.73(1.06) &  483.99 & 3.784(4) &       \\
     &  68 &  591.27(56)   &  590.41 & 3.901(4) &       \\
  Zr &  90 &  783.35(46)   &  783.89 & 4.263(3) & 4.269 \\
  Sn & 100 &  812.91(1.18) &  824.79 & 4.480(5) &      \\
     & 132 & 1100.80(54)   & 1102.85 & 4.714(4) & 4.709 \\
  Pb & 208 & 1637.81(67)   & 1636.43 & 5.479(5) & 5.501 \\
  \end{tabular}
  \end{ruledtabular}
\end{table}

\begin{figure}
\includegraphics[width=8cm]{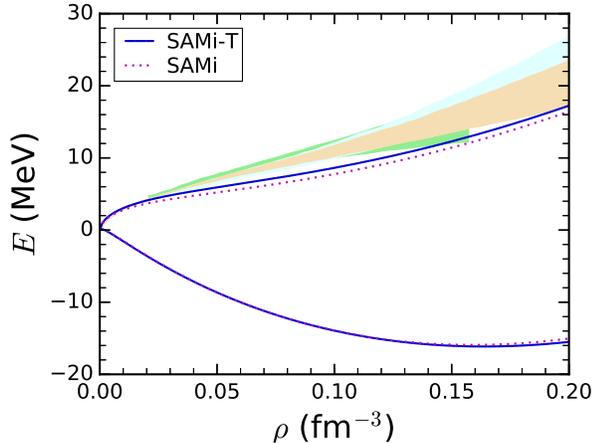}
\caption{(Color online) Neutron matter and symmetric nuclear matter equation of state calculated by new fitted Skyrme functional SAMi-T (blue solid line), in comparison with those of SAMi functional (purple dotted line) \cite{Roca-Maza2012}. The shaded area are constraints of neutron matter from different \textit{ab initio} calculations, see text for detail.}
\label{fig:eos}
\end{figure}

Figure~\ref{fig:eos} shows the Equation of State (EoS) for both neutron matter and symmetric nuclear matter calculated by the new fitted Skyrme functional SAMi-T, in comparison with those of SAMi functional. For the neutron matter, we also show the results of different \textit{ab initio} calculations: self-consistent Green's function method with chiral $NN$ and $3N$ forces up to N$^3$LO \cite{Drischler2016} (cyan shade); quantum Monte-Carlo method with chiral $NN$ and $3N$ forces up to N$^2$LO \cite{Lynn2016} (green shade); loop expansion around the Hartree-Fock energy with chiral $NN$ force up to N$^3$LO and $3N$ force up to N$^2$LO \cite{Hebeler2010a} (brown shade), see also Figure 1 from Ref.~\cite{Tews2017}.

Since the fitting protocol of SAMi-T is similar to that of SAMi, the symmetric matter equation-of-state of the two functionals are almost identical. On the other hand, the symmetry energy at saturation density of SAMi-T, 
 $J = 29.7\pm 0.6$ MeV, is slightly larger than that of SAMi, $J_{\rm SAMi} = 28\pm 1$ MeV. Therefore, the equation-of-state of the neutron matter given by SAMi-T shows more repulsion, and this is also in better agreement with the constraints from different \textit{ab initio} calculations. The slope of the symmetry energy at saturation density of SAMi-T is $L = 46\pm 12$ MeV, also slightly larger than that of SAMi, $L_{\rm SAMi} = 44\pm 7$ MeV.

\begin{figure}
\includegraphics[width=8cm]{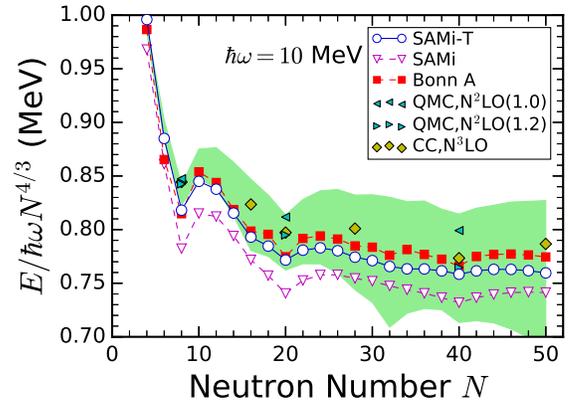}
\caption{(Color online) Total energy (in units of $\hbar\omega N^{4/3}$) of neutron drops with $N$ from $8$ to $50$ in a HO trap ($\hbar\omega = 10$ MeV) calculated by SAMi-T, in comparison with results of SAMi and of RBHF theory using the Bonn A interaction \cite{Shen2018}, Quantum Monte-Carlo method using local chiral interaction N$^2$LO with two cut-offs $1.0$ fm and $1.2$ fm \cite{Tews2016}, coupled-cluster theory using chiral interaction N$^3$LO \cite{Potter2014}.
The shaded area represents the values spanned by quantum Monte-Carlo calculations using AV8'+UIX and AV8'+IL7 \cite{Gandolfi2011,Maris2013}.}
\label{fig:ea-nd}
\end{figure}

In Fig.~\ref{fig:ea-nd}, we show the total energy (in units of $\hbar\omega N^{4/3}$) of neutron drops from $N = 8$ to $50$ in a HO trap ($\hbar\omega = 10$ MeV) calculated by SAMi-T, in comparison with results of SAMi, RBHF calculations using the Bonn A interaction \cite{Shen2018}, and QMC 
calculations using AV8'+UIX (upper bound of the shaded area) and AV8'+IL7 (lower bound of the shaded area) \cite{Gandolfi2011,Maris2013}, QMC using local chiral interaction N$^2$LO with two cut-offs $1.0$ fm and $1.2$ fm \cite{Tews2016}, coupled-cluster theory using chiral interaction N$^3$LO \cite{Potter2014}.
The no-core shell model using same chiral interaction N$^3$LO (calculated up to $N = 18$) coincide with those by coupled-cluster theory \cite{Potter2014} and will not be plotted.
As it can be expected from the EoS in Fig.\ref{fig:eos}, the total energy of neutron drops given by SAMi-T are generally larger than those of SAMi, but smaller than those of RBHF with Bonn A calculations.
The error we adopted for the energy of neutron drops in the fitting, $\Delta O_i = 0.01$ for $E/(\hbar\omega N^{4/3})$, assumes a clearer meaning by looking at Fig.~\ref{fig:ea-nd}: in fact, 
the values differing from the results of RBHF with Bonn A by about 0.01 are seen to be acceptable for our purpose.

Comparing with other (nonrelativistic) \emph{ab initio} calculations using chiral interactions, the energy given by SAMi-T (or those by RBHF with Bonn A) are slightly lower, similar as in the comparison in Fig.~\ref{fig:eos}.

\begin{figure}
\includegraphics[width=8cm]{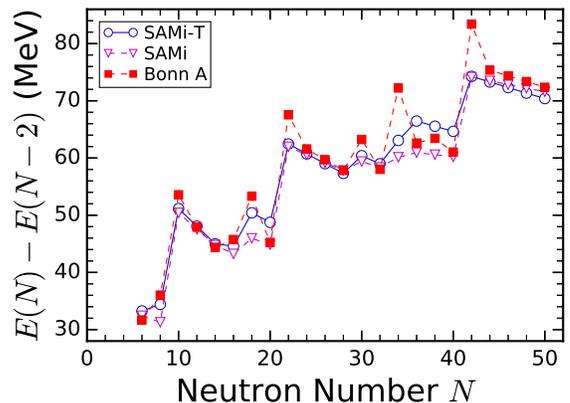}
\caption{(Color online) Two-neutron energy difference of neutron drops in a HO trap ($\hbar\omega = 10$ MeV) calculated by SAMi-T, in comparison with results of SAMi and of the RBHF theory 
using the Bonn A interaction \cite{Shen2018}.}
\label{fig:s2n-nd}
\end{figure}

To see the subshell structure in neutron drops, we show in Fig.~\ref{fig:s2n-nd} the two-neutron separation energies $E(N)-E(N-2)$ for the same calculations we have just mentioned. Since those calculations are in a HO external field, they all show clear shell structure for $N = 8, 20$, and $40$. Besides those, the subshell structure at $N = 16, 32$, and a somewhat weaker one $N = 28$ can also be seen. For these subshells, SAMi-T also shows similar trends as the RBHF calculations. Here the results of both functionals is very similar, with some improvement of SAMi-T for $N=18$ and somehow also in the region $34<N<38$.
The different peak position in this region, for RBHF at $N = 34$ (and a minor one at $N = 38$) and for SAMi-T (or SAMi) at $N = 36$ is due to different ordering and different occupations of the single-particle levels:
For RBHF, from $N = 32$ to $N = 34$, the $\nu 1f_{5/2}$ level is being occupied and the energy suddenly increases much; from $N = 34$ to $N = 36$ the $\nu 1f_{5/2}$ level continues to be occupied and the energy increases less; from $N = 36$ to $N = 38$, the level $\nu 2p_{1/2}$ is being occupied before the level $\nu 1f_{5/2}$ is filled at $N = 40$, therefore another small jump appears at $N = 38$.
For SAMi-T (or SAMi) functional, from $N = 32$ to $N = 34$ the level $\nu 2p_{1/2}$ is firstly filled and from $N = 34$ to $N = 36$ the $\nu 1f_{5/2}$ begins to be occupied.
Therefore, the jump of the energy is smooth and the two-neutron separation energy keeps increasing from $N = 32$ to $N = 36$.
From $N = 36$ to $N = 40$ the level $\nu 1f_{5/2}$ continues to be filled until fully occupied, the energy increases less and therefore the $S_{2n}$ decreases smoothly.

\begin{figure}
\includegraphics[width=8cm]{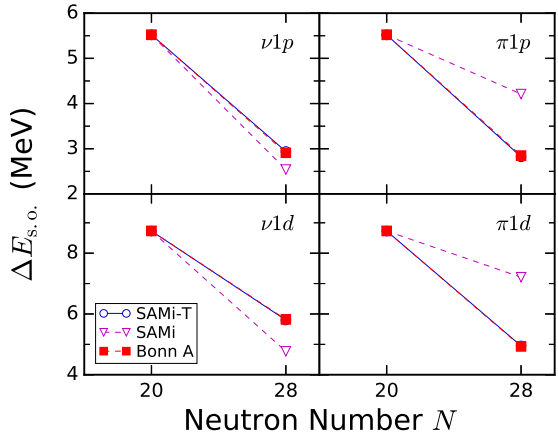}
\caption{(Color online) Neutron and proton $1p$ and $1d$ spin-orbit splittings of neutron-proton drops calculated by SAMi-T, in comparison with results of SAMi functional and RBHF theory using the Bonn A interaction. The results of SAMi-T and SAMi have been shifted so that the first points ($N = 20$) meet with those of RBHF.}
\label{fig:so-npd}
\end{figure}

Figure~\ref{fig:so-npd} shows the neutron and proton $1p$ and $1d$ spin-orbit splittings of neutron-proton drops, whose trends are used as pseudodata in the fitting procedure and are mainly responsible for constraining the tensor force.
Again, we show the results given by SAMi-T, and comparing with those of SAMi functional and the pseudodata of RBHF with Bonn A interaction.
The results of SAMi-T and SAMi have been shifted so that the first points ($N = 20$) meet with those of RBHF.
The shifts for SAMi-T are $\Delta E_{\rm s.o.}^{\rm (SAMi-T)} - \Delta E_{\rm s.o.}^{\rm (RBHF)} = 1.03$ MeV ($1p$) and $2.43$ MeV ($1d$); while for SAMi they are $0.03$ MeV ($1p$) and $0.86$ MeV ($1d$).
This is expected as the absolute value of spin-orbit splittings of SAMi-T are fitted to experimental value and they are larger than those of RBHF, as explained in the numerical details.
On the other hand, SAMi-T reproduces the relative changes, which reflects the effect of tensor force, very well.
For the relative change of spin-orbit splittings, the results of SAMi-T are exactly on top of those of RBHF.

In Refs.~\cite{Otsuka2005,Shen2018}, these trends for the evolution of the spin-orbit splitting are explained by the effect of tensor forces. For the spin-saturated system with $N = 20$, there is no tensor force contribution from the neutrons; for the spin-unsaturated system with $N = 28$, where only one of the spin-orbit partners is occupied ($1f_{7/2}$) while the other is empty ($1f_{5/2}$), the tensor force reduces the spin-orbit splittings dramatically. This is shown in the RBHF calculation with Bonn A in Fig.~\ref{fig:so-npd}, where the left panels depict the tensor force effect in the neutron-neutron channel and the right panels correspond to effect in the neutron-proton channel.
For pure neutron drops within the RBHF calculation, a similar phenomenon has been shown in Ref.~\cite{Shen2018}. 

In the Skyrme functional, the tensor force in the same particle channel (neutron-neutron or proton-proton) is controlled by the parameter $\alpha$ and the tensor force in the different particle channel is controlled by $\beta$ in Eq.~(\ref{eq:Uso}).
Both $\alpha$ and $\beta$ have two origins, one is from the exchange terms of the central interaction 
($\alpha_{c}$ and $\beta_c$) and another one is from the tensor force ($\alpha_{T}$ and $\beta_T$) as is 
clearly shown in Eq.~(\ref{eq:alpha-beta}).
In Table~\ref{tab:alpha-beta}, we list the values of $\alpha$ and $\beta$ for SAMi-T and SAMi functionals.
Without tensor force, the contribution to $\alpha$ and $\beta$ in SAMi functional comes only 
from the central part.
One finds that the value of $\alpha$ in SAMi is slightly larger than that of SAMi-T, and therefore SAMi 
shows similar trends but with larger magnitude in the left panels of Fig.~\ref{fig:so-npd}, where 
only the neutron-neutron channel matters.
With only the contribution from the central force, which is also very much determined 
by properties like binding energies and charge radii, SAMi has less freedom to fit to the 
spin-orbit splittings in Fig.~\ref{fig:so-npd}.
As a consequence, the value of $\beta$ in SAMi 
is much smaller than that of SAMi-T, and SAMi does not show the trends of the right panels 
of Fig.~\ref{fig:so-npd}, where the neutron-proton channel matters.

In conclusion, the tensor terms in the new Skyrme functional are determined reasonably and,
more importantly, without ambiguities. Had we put more emphasis on the evolution of the 
experimental single-particle states, we would have faced the problem that the coupling with 
collective vibrations may play an important role there \cite{Litvinova2011,Afanasjev2015a}. 
Instead, fitting to the pseudodata of \textit{ab initio} RBHF calculations does not imply this ambiguity.

\begin{table}[!th]
  \caption{Values of $\alpha$ and $\beta$ in Eq.~(\ref{eq:Uso}) of SAMi-T (with estimated standard deviations in the parentheses) and SAMi \cite{Roca-Maza2012}.}
  \label{tab:alpha-beta}
  \centering
  \begin{ruledtabular}
  \begin{tabular}{lrr}
  & $\alpha$ (MeV fm$^5$) & $\beta$ (MeV fm$^5$) \\
  \hline
  SAMi-T  & $73.0(8)$ & $101.8(1.2)$ \\
  SAMi \cite{Roca-Maza2012} & $101.6$ & $31.5$ \\
  \end{tabular}
  \end{ruledtabular}
\end{table}

As a side remark, we note that 
from Fig.~\ref{fig:so-npd} we can also understand why SAMi gives a smaller value for the 
spin-orbit strength because of the lack of tensor terms: both the new functional and SAMi have been 
fit using the proton spin-orbit splittings in neutron spin-unsaturated system 
($^{90}$Zr with $N = 50$, where the $1g_{9/2}$ orbit is occupied and the $1g_{7/2}$ 
orbit is empty; $^{208}$Pb with $N=126$, where the $1i_{13/2}$ orbit is occupied and the 
$1i_{11/2}$ orbit is empty), so that the functional with tensor terms must have a stronger 
spin-orbit strength $W_0$ (or $W_0'$) as the inclusion of the tensor force 
will imply a decrease of the SO splittings.

\begin{figure}
\includegraphics[width=8cm]{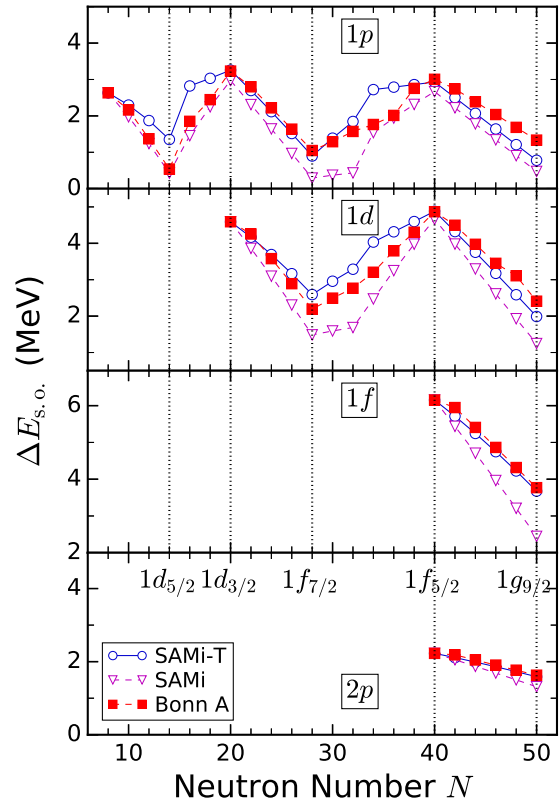}
\caption{(Color online) From top to bottom panel, $1p, 1d, 1f$, and $2p$ spin-orbit splittings of $N$-neutron drops in a HO trap ($\hbar\omega = 10$ MeV) calculated by SAMi-T, in comparison with results of SAMi functional and RBHF theory using the Bonn A interaction \cite{Shen2018}.}
\label{fig:so-nd}
\end{figure}

Since the values of $\alpha$ are similar in SAMi-T and SAMi, we can expect the two functionals show similar trends in the evolution of spin-orbit splittings in neutron drops, and this is shown in Fig.~\ref{fig:so-nd}.
Similarly as in Fig.~\ref{fig:so-npd}, the results of SAMi-T and SAMi have been shifted so that the first points meet with those of RBHF: in other words, we put emphasis here on the relative changes rather
than the absolute values.
The shifts for SAMi-T are $\Delta E_{\rm s.o.}^{\rm (SAMi-T)} - \Delta E_{\rm s.o.}^{\rm (RBHF)} = 1.32$ MeV ($1p$), $2.28$ MeV ($1d$), $3.30$ MeV ($1f$), $1.25$ MeV ($2p$); while for SAMi they are $0.60$ MeV ($1p$), $0.88$ MeV ($1d$), $1.28$ MeV ($1f$), $0.52$ MeV ($2p$).
As mentioned in the numerical settings, the pseudodata of spin-orbit splittings by RBHF are used only to constrain the tensor terms $\alpha_T$ and $\beta_T$, but not the absolute strength of spin-orbit terms. The spin-orbit term have been constrained by the experimental spin-orbit splittings.

Also in this case, SAMi-T shows better agreement with the results of RBHF calculations.
With fitting to the trend of the spin-orbit splitting in neutron-proton drops given by RBHF (as 
shown in Fig.~\ref{fig:so-npd}), the trend of the same splitting in neutron drops shown in 
Fig.~\ref{fig:so-nd} can be reproduced automatically.
It should also be noted that in detail, certain deviations exist.
Using the results of RBHF as a baseline, the decrease of spin-orbit splittings by SAMi-T is smaller around $N = 14$ ($1p$), and getting stronger around $N = 28$ ($1p$ slightly larger while $1d$ slightly smaller), and even more around $N = 50$ ($1p, 1d$ slightly larger while $1f, 2p$ agree).
Similar trend is found for SAMi.
This may be understandable considering the fact that the tensor terms in the Bonn A interaction are finite-range one-boson exchange interactions, while the tensor terms in Skyrme functional (\ref{eq:vt}) are zero-range $\delta$ interactions.
Another cause may be that the difference of the density distributions can also significantly influence the change of spin-orbit splittings.
Overall, we do not expect an exact matching with RBHF results but just to achieve the best possible fit.

\begin{figure}
\includegraphics[width=8cm]{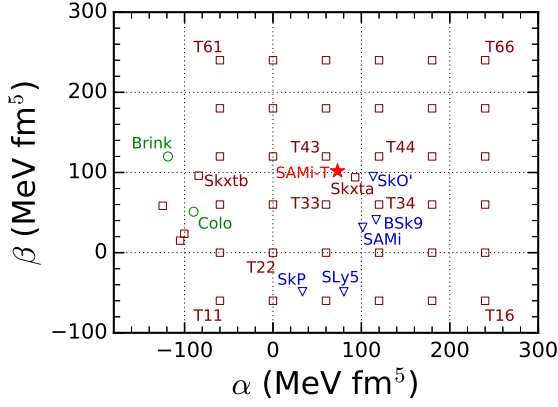}
\caption{(Color online) 
Values of $\alpha$ and $\beta$ in Eq.~(\ref{eq:Uso}) from different Skyrme functionals. See text for the detail.}
\label{fig:ab}
\end{figure}

In Fig.~\ref{fig:ab}, we show the values of $\alpha$ and $\beta$ from different Skyrme functionals, including three categories:
\begin{enumerate}
  \item Skyrme functionals that do not contain tensor terms $\alpha_T$ and $\beta_T$ but take the $\mathbf{J}^2$ from the central force into account (blue triangles in the figure).
  Those are SkP \cite{Dobaczewski1984}, SLy5 \cite{Chabanat1998}, SkO' \cite{Reinhard1999}, BSk9 \cite{Goriely2005}, SAMi \cite{Roca-Maza2012}; 
  \item Skyrme functionals in which the tensor terms are added perturbatively without refitting the parameters (green circles in the figure), such as SLy5+T by Col\`o \textit{et al.} \cite{Colo2007} or SIII+T by Brink and Stancu \cite{Brink2007}.
  \item Skyrme functionals in which the tensor terms are included on equal footing with other terms and an overall fit is performed (brown squares in the figure), such as Skxta and Skxtb \cite{Brown2006}, the TIJ family (I,J from $1$ to $6$) \cite{Lesinski2007}, SkP-T, SLy4-T and SkO-T \cite{Zalewski2008}, and the present SAMi-T (red star in the figure).
\end{enumerate}

Among different Skyrme functionals, the $\alpha, \beta$ values of SAMi-T are close to those of T43, Skxta, and SkO'.
The tensor part of Skxta is determined by the nonrelativistic HKT $G$-matrix \cite{Hosaka1985}, which reproduces the $G$-matrix elements obtained from the bare nucleon-nucleon Paris interaction \cite{Lacombe1980}.
It is interesting to see that the tensor force given by the relativistic $G$-matrix of Bonn A interaction is similar to the nonrelativistic $G$-matrix of Paris interaction.
For SkO', the tensor force is not included and therefore all the contributions to $\alpha$ and $\beta$ come from the central part $\alpha_c,\beta_c$.
This place a tight constraint for the central force and the ground-state properties are sacrificed to some extent.
For example, the binding energy of $^{208}$Pb given by SkO' is $B_{\rm 208Pb}^{\rm SkO'} = 1644.09$ MeV, while the experimental data is $1636.43$ MeV and the one given by SAMi-T is $B_{\rm 208Pb}^{\rm SAMi-T} = 1637.81\pm 0.67$ MeV (see Table \ref{tab:erso-fn}).

\subsection{Excited states}

In this Subsection, we investigate some of the excited state properties within the self-consistent HF plus Random Phase Approximation (RPA) \cite{Colo2013} using the new fitted Skyrme functional SAMi-T.
The basis for the configuration space RPA calculations is chosen by including all occupied states, and unoccupied states with an angular momentum cut-off $l_{\rm cut} = 10\hbar$, and $10$ states with increasing values of the $n$-quantum number in each $(l,j)$ block.

\begin{figure}
\includegraphics[width=8cm]{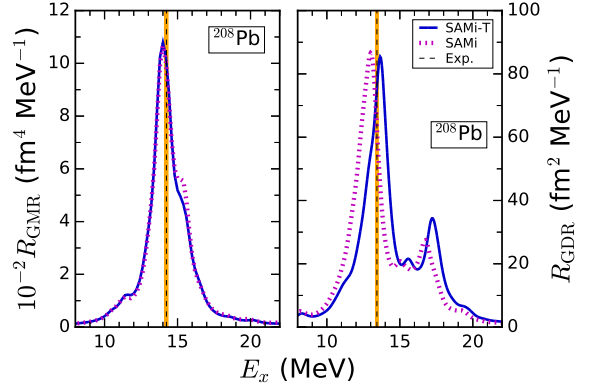}
\caption{(Color online) Strength function associated with the GMR (left) and IVGDR (right), calculated 
by using the SAMi-T functional and compared with results of SAMi \cite{Roca-Maza2012} and with the experimental centroid energies \cite{Youngblood1999,Ryezayeva2002}.
For theoretical results, we adopt a smoothing with Lorentzian functions having $1$ MeV width.}
\label{fig:gmr}
\end{figure}

First of all, we would like to confirm that SAMi-T does perform as well as other functionals for non-charge exchange resonances. With this aim, we show in Figure~\ref{fig:gmr}, as representative examples, the strength function $R(E)$ associated with the isoscalar giant monopole resonance (GMR) and the isovector giant dipole resonance (IVGDR): the results obtained with SAMi-T are compared with results from the SAMi functional \cite{Roca-Maza2012} and with the experimental centroid energies \cite{Youngblood1999,Ryezayeva2002}.
The strength function is defined as \cite{Colo2013}
\begin{equation}\label{eq:}
  R(E) \equiv \sum_{\nu} |\langle \nu||\hat{F}_J||0\rangle|^2 \delta(E-E_{\nu}),
\end{equation}
  where $\hat{F}_J$ is the multipole excitation operator with total angular momentum $J$, $|0\rangle$ is the ground state, and $|\nu\rangle$ is the excited state with energy $E_{\nu}$.
The operators used in GMR and IVGDR calculations are
\begin{equation}\label{eq:}
  \hat{O}_{\rm GMR} = \sum_{i=1}^A r_i^2, \quad \hat{O}_{\rm IVGDR} = \frac{Z}{A}\sum_{n=1}^N r_n - \frac{N}{A}\sum_{p=1}^Z r_p,
\end{equation}
respectively.

The centroid energy of GMR by SAMi-T is $E_c^{\rm SAMi-T}({\rm GMR}) = 14.4\pm 0.2$ MeV, in good agreement with the experimental value $E_c^{\rm exp}({\rm GMR}) = 14.24\pm 0.11$ MeV; our calculation 
exhausts $98.4\%$ of the energy weighted sum rule (EWSR) between $E = 8$ and $22$ MeV.
The centroid energy of IVGDR by SAMi-T is $E_c^{\rm SAMi-T}({\rm IVGDR}) = 14.6\pm1.3$ MeV, slightly larger than the experimental value $E_c^{\rm exp}({\rm IVGDR}) = 13.43\pm 0.10$ MeV; we fulfil $94.9\%$ of the EWSR between $E = 9$ and $20$ MeV.
We have also calculated using SAMi-T without tensor and the results are very similar to SAMi-T with tensor, therefore we do not plot them in the figure.
In this subsection, unless specified otherwise, all calculations without tensor refer to dropping the 
tensor in the residual interaction of the RPA calculation, but the tensor 
is kept in Hartree-Fock.

\begin{figure}
\includegraphics[width=8cm]{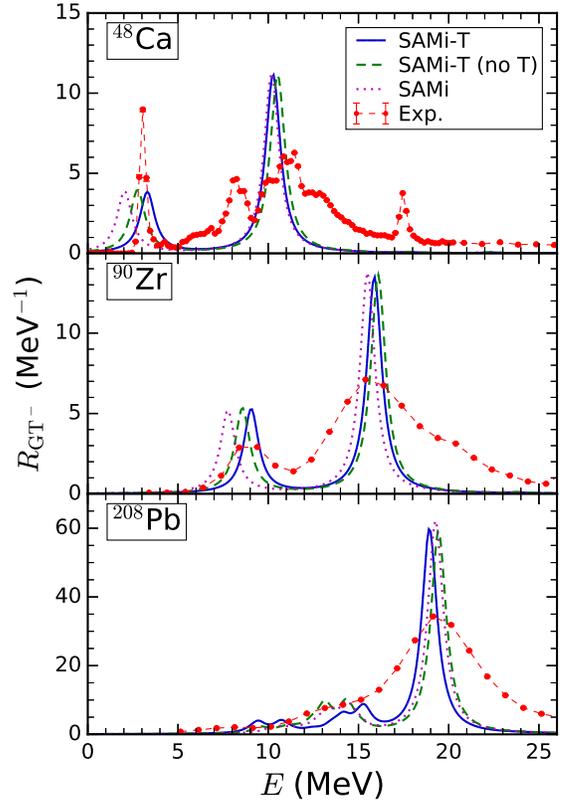}
\caption{(Color online) GTR strength function for $^{48}$Ca, $^{90}$Zr, and $^{208}$Pb calculated by SAMi-T with and without tensor, in comparison with experimental data \cite{Wakasa1997,Yako2009,Krasznahorkay2001,Akimune1995,Wakasa2012} and with the results of the SAMi functional \cite{Roca-Maza2012}.
For theoretical results, we adopt a smoothing with Lorentzian functions having $1$ MeV width.}
\label{fig:gt}
\end{figure}

We now study two different types of nuclear collective excitation where the spin and spin-isospin channels of the functional should play a crucial role. 
In Fig.~\ref{fig:gt}, we show the GTR strength function in $^{48}$Ca, $^{90}$Zr, and $^{208}$Pb calculated by SAMi-T with and without tensor, in comparison with experimental data \cite{Wakasa1997,Yako2009,Krasznahorkay2001,Akimune1995,Wakasa2012} and results of SAMi functional \cite{Roca-Maza2012}.
The operators for the GTR calculation is
\begin{equation}\label{eq:}
  \hat{O}_{\rm GTR} = \sum_{i=1}^A \bm{\sigma}(i) \tau_{\pm}(i),
\end{equation}
where $\bm{\sigma}$ is the spin, $\tau_+ (\tau_-)$ is the isospin raising (lowering) operator.
In all cases, the non-energy weighted sum rule (NEWSR)
\begin{equation}\label{eq:}
\int \left[ R_{GT^-}(E)-R_{GT^+}(E)\right] dE = 3(N-Z),
\end{equation}
is fulfilled by the theoretical calculations.

For $^{48}$Ca, the locations of low-energy peak and high-energy peak given by SAMi-T are $E_{\rm low}^{\rm SAMi-T} = 3.3\pm1.3$ MeV and $E_{\rm high}^{\rm SAMi-T} = 10.3\pm0.3$ MeV, and they are in good agreement with the experimental values: $E_{\rm low}^{\rm exp} = 3.0$ MeV and $E_{\rm high}^{\rm exp} = 10.5$ MeV.
For $^{90}$Zr, the locations given by SAMi-T are $E_{\rm low}^{\rm SAMi-T} = 9.1\pm1.0$ MeV and $E_{\rm high}^{\rm SAMi-T} = 15.9\pm0.2$ MeV, and they also agree well with the experimental values: $E_{\rm low}^{\rm exp} = 9.0\pm 0.5$ MeV and $E_{\rm high}^{\rm exp} = 15.8\pm 0.5$ MeV.
In comparison with SAMi, the high-energy peaks in these two nuclei described by the new functional are similar, and the low-energy peaks have been improved.
For $^{208}$Pb, the peak location given by SAMi-T is $E^{\rm SAMi-T} = 19.0\pm0.3$ MeV, also in good agreement with the experimental value $E^{\rm exp} = 19.2\pm 0.2$ MeV.

In the GTR shown in Fig.~\ref{fig:gt}, the tensor force shows some influence but not a huge one.
The effects of tensor force, however, are more significant in the case of the SDR.
In Ref.~\cite{Bai2010}, it has been found that tensor correlations have a unique, multipole-dependent effect on the three SDR components: the $1^-$ state is being pushed to higher energy, while the $0^-$ and $2^-$ states are being pushed to lower energy.
Quantitatively, the residual interaction matrix element in the $0^-$ is the largest, the matrix element for $1^-$ is the next largest, and the effect on $2^-$ is rather small \cite{Bai2010}.

\begin{figure}
\includegraphics[width=8cm]{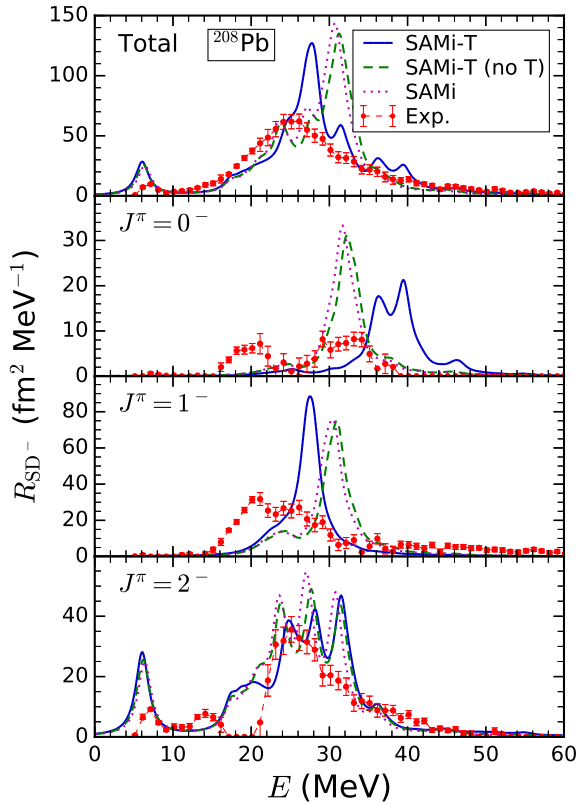}
\caption{(Color online) SDR strength function in the $\tau_-$ channel for $^{208}$Pb, calculated by SAMi-T with and without tensor, in comparison with experimental data \cite{Wakasa2012} and the results with SAMi \cite{Roca-Maza2012}.
From top to bottom, the total strength and the $J^\pi = 0^-, 1^-$, and $2^-$ components are shown.
For theoretical results, we adopt a smoothing with Lorentzian functions having $2$ MeV width.}
\label{fig:sd}
\end{figure}

In Fig.~\ref{fig:sd}, we show the SDR strength function in the $\tau_-$ channel for $^{208}$Pb ($J^\pi = 0^-, 1^-, 2^-$ and total). We compare, once again, the results obtained by SAMi-T with and without tensor 
with experimental data \cite{Wakasa2012} and the results with the SAMi functional \cite{Roca-Maza2012}.
The operator used for the RPA calculations is
\begin{equation}\label{eq:}
  \hat{O}_{\rm SDR} = \sum_{i=1}^A \sum_M \tau_{\pm}(i) r_i^L [Y_L(\hat{r}_i)\otimes \bm{\sigma}(i)]_{JM}.
\end{equation}

Without tensor force, the new functional gives similar results as SAMi.
It can be seen that the effect of tensor force is consistent with what has been found in Ref.~\cite{Bai2010}, namely different channels show different effects:
For the $J^\pi = 1^-$, the tensor force improves the description for the data while for $J^\pi = 0^-$ does not, and for $J^\pi = 2^-$ the influence of the tensor force is very small.
As the $1^-$ channel gives the largest contribution to the total SDR, the overall agreement with the experimental data is improved by including the tensor force.
The experimental (SAMi-T) NEWSR is $107_{-7}^{+8}$ ($160\pm1$ fm$^2$) for $J^\pi = 0^-$, $450_{-15}^{+16}$ ($441\pm3$ fm$^2$) for $J^\pi = 1^-$, $447_{-15}^{+16}$ ($660\pm6$ fm$^2$) for $J^\pi = 2^-$, and $1004_{-23}^{+24}$ fm$^2$ ($1260\pm10$ fm$^2$) for the total.
The NEWSR given by SAMi-T is similar compared with SAMi \cite{Roca-Maza2012}.

The sum rule,
\begin{equation}\label{eq:}
  \int \left[ R_{SD^-}(E)-R_{SD^+}(E)\right] dE = \frac{9}{4\pi}(N\langle r_n^2\rangle - Z \langle r_p^2\rangle ),
\end{equation}
is fully exhausted in the calculation.

As shown by the above formula, the sum rule of SDR is related with the neutron and proton rms radius.
Since the sum rule given by SAMi-T can well reproduce the experimental data, and the proton (charge) radius of $^{208}$Pb is well fitted by SAMi-T (see Table~\ref{tab:erso-fn}), it would be interesting to show the neutron skin too.
The neutron skin of $^{208}$Pb calculated by SAMi-T is $\Delta r_{\rm np} = \langle r_n^2\rangle^{1/2} - \langle r_p^2\rangle^{1/2} = 0.153\pm0.011$ fm, to be compared with some recent experimental data $0.33^{+0.16}_{-0.18}$ fm (from parity violation \cite{prex}), $0.16\pm 0.06$ fm (from measurements with antiprotonic atoms \cite{Klos2007}), $0.15^{+0.04}_{-0.06}$ fm (pion photoproduction \cite{Tarbert2014}), $0.165\pm0.043$ fm (electric dipole polarizability \cite{Roca-Maza2013}).

In Refs.~\cite{Bai2010,Bai2011a}, different strengths of tensor terms have been proposed, on top of 
the SLy5 functional \cite{Chabanat1998}. In connection with the SDR study, it was found
in order to improve the $1^-$ strength function while not make the $0^-$ strength function worse, a relatively large value of $\alpha$ and $\beta$ is needed (in Ref.~\cite{Bai2010} $\alpha = 217$ MeV fm$^5$ and $\beta = 189$ MeV were suggested).
Guided by the calculations from RBHF theory with Bonn A interaction, such values for the tensor terms are too strong.
How to improve the description of the $0^-$ channel in ${}^{208}$Pb while not enter in contradiction with \textit{ab initio} results is an interesting question for future investigations.

\begin{figure}
\includegraphics[width=8cm]{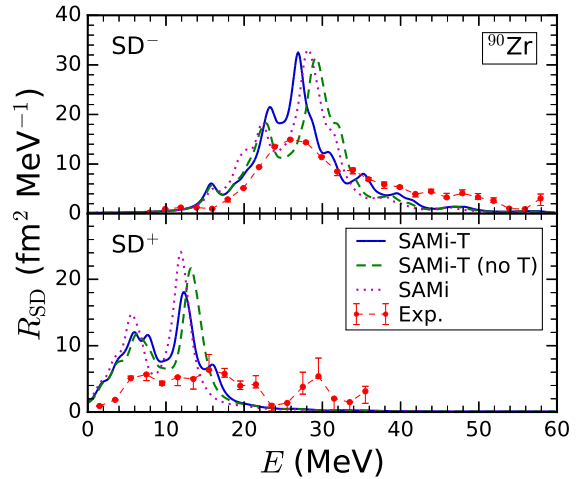}
\caption{(Color online) SDR strength function of the $\tau_-$ (upper) and $\tau_+$ (lower) channel in $^{90}$Zr calculated by SAMi-T with and without tensor, in comparison with experimental data \cite{Yako2006} and SAMi functional \cite{Roca-Maza2012}.
For theoretical results, a Lorentzian smearing parameter equal to $2$ MeV is used.}
\label{fig:sd-zr}
\end{figure}

Finally, in Fig.~\ref{fig:sd-zr}, we show the SDR strength function of the $\tau_-$ and $\tau_+$ channel in $^{90}$Zr calculated by SAMi-T with and without tensor, in comparison with experimental data \cite{Yako2006} and SAMi functional \cite{Roca-Maza2012}.
The peak position of $R_{\rm SD^-}$ has been improved by including the tensor, while for $R_{\rm SD^+}$ the effect is not significant.
The NEWSR calculated by the new functional is $147\pm1$ fm$^2$ ($100\%$ exhausted in the calculation), while the experimental value is $148\pm 12$ fm$^2$.
The neutron skin of $^{90}$Zr calculated by SAMi-T is $\Delta r_{\rm np} = 0.069\pm0.005$ fm, to be compared with experimental data $0.07\pm 0.04$ fm \cite{Yako2006}.

\section{Summary}\label{sec:sum}

In this work we have developed a new Skyrme functional SAMi-T inspired by the fitting protocol of the successful SAMi functional, with further information on tensor terms provided by \textit{ab initio} calculations. For that purpose we have calculated for the first time neutron-proton drops as predicted by RBHF calculations based on the Bonn A potential.
With the evolution of SO splittings as a function of neutron number in the neutron-proton drops the tensor terms are well constrained.
The new Skyrme functional is then used to investigate the ground- and excited-state properties of nuclei.

For the ground state properties of spherical nuclei, such as binding energy and charge radius, the new functional SAMi-T can achieve similar precision as other successful functionals.
Therefore, the introduction of strong tensor terms in the Skyrme functional does not decrease the accuracy of describing ground state properties.

By fitting only to the evolution of SO splittings in the neutron-proton drops as predicted by RBHF theory, the evolution in the neutron drops by RBHF can be reproduced automatically by the new functional.
In other words, the information of the tensor in the neutron-neutron channel is consistent in these two systems. 

The new functional SAMi-T also gives a good description of the excitation energies and sum rules of Giant Resonances. We have investigated the giant-monopole resonance and giant-dipole resonance of $^{208}$Pb; the Gamow-Teller resonance of $^{48}$Ca, $^{90}$Zr, and $^{208}$Pb; the spin-dipole resonance of $^{90}$Zr and $^{208}$Pb.

The new functional, as in the case of SAMi, respects the empirical hierarchy of the spin and spin-isospin Landau-Migdal parameters $G_0' > G_0 > 0$. This has two advantages: i) pushes the model to produce stable results in infinite matter around saturation \cite{Pastore2015}.
The instability of SAMi-T is briefly studied and presented in Appendix \ref{app:stable};
ii) ensures a better reproduction of spin and spin-isospin resonances.
For the GTR of $^{48}$Ca and $^{90}$Zr, the lower-energy and higher-energy peak are both reproduced very well and, to our knowledge, no other interaction of the same type is as accurate as SAMi-T in these two cases.
The higher-energy peaks of these two cases are well described by both SAMi and SAMi-T, but the lower-energy peaks are improved by SAMi-T.
For the GTR of $^{208}$Pb, the peak position is also nicely reproduced by SAMi-T, similar as SAMi.

The SDR components are also improved comparing with SAMi functional, especially in the case of the $1^-$ channel in $^{208}$Pb which is improved directly by the tensor term.
As the NEWSR of SDR is related with the neutron skin, and for $^{208}$Pb and $^{90}$Zr they can be well described by SAMi-T, the predictions of neutron skin of these two nuclei by SAMi-T are also in good agreement with experimental data.

\section*{ACKNOWLEDGMENTS}

We would like to thank L. G. Cao for cross checking the results; D. Davesne and A. Pastore for discussions and providing results of instability analysis; J. Meng and H. Sagawa for stimulating discussions.
This work was partly supported by Funding from the European Union's Horizon 2020 research and innovation programme under Grant agreement No. 654002.

\appendix

\section{Covariance analysis}\label{app:err}

In the fit of the new Skyrme functional SAMi-T, the $\chi^2$ is defined as
\begin{equation}\label{eq:}
  \chi^2(\mathbf{p}) = \sum_{i=1}^m \left( \frac{\mathcal{O}_i^{\rm theo.}(\mathbf{p})-\mathcal{O}_i^{\rm ref.}}{\Delta \mathcal{O}_i^{\rm ref.}} \right)^2,
\end{equation}
where $\mathbf{p} = (p_1,\dots,p_n)$ is the $n$ parameters, $\mathcal{O}_i$ are observables with $m$ data points, `theo.' stands for the calculated values and `ref.' for experimental data or, in this work, pseudodata from RBHF calculations.
The corresponding adopted errors are denoted as $\Delta \mathcal{O}_i^{\rm ref.}$.
See Table.~\ref{tab:datafit} for the detailed information in the fit of SAMi-T.

Assuming the $\chi^2$ reaches a minimum and is a well behaved function of the parameters around the optimal value $\mathbf{p}_0$, it can be expanded and approximated as
\begin{equation}\label{eq:}
  \chi^2(\mathbf{p}) - \chi^2(\mathbf{p}_0) \approx \frac{1}{2} \sum_{i,j}^n (p_i-p_{0,i}) \partial_{p_i}
  \partial_{p_j} \chi^2 \big|_{\mathbf{p}_0}(p_j-p_{0,j}).
\end{equation}
The curvature matrix is defined as
\begin{equation}\label{eq:}
  \mathcal{M}_{ij} \equiv \partial_{p_i}\partial_{p_j} \chi^2 \big|_{\mathbf{p}_0}.
\end{equation}
The error (or covariance) matrix is defined as the inverse of the curvature matrix,
\begin{equation}\label{eq:}
  \mathcal{E} = \mathcal{M}^{-1}.
\end{equation}
The error for parameter $p_i$ can be obtained as
\begin{equation}\label{eq:}
  e(p_i) = \sqrt{\mathcal{E}_{ii}}.
\end{equation}
The correlation between different parameters can be evaluated by the correlation matrix
\begin{equation}\label{eq:}
  \mathcal{C}_{ij} \equiv \frac{\mathcal{E}_{ij}}{\sqrt{\mathcal{E}_{ii}\mathcal{E}_{jj}}}.
\end{equation}

Assuming the observable $A(\mathbf{p})$ are smooth functions around the optimal value $\mathbf{p}_0$, one has
\begin{equation}\label{eq:}
  A(\mathbf{p}) \approx A(\mathbf{p}_0) + (\mathbf{p}-\mathbf{p}_0) \partial_{\mathbf{p}} A(\mathbf{p})|_{\mathbf{p}=\mathbf{p}_0}.
\end{equation}
Within this approximation, one can calculate the covariance between two observables $A(\mathbf{p})$ and $B(\mathbf{p})$ as
\begin{equation}\label{eq:}
  C_{AB} = \sum_{ij}^n \frac{\partial A(\mathbf{p})}{\partial p_i}\bigg|_{\mathbf{p}=\mathbf{p}_0}
  \mathcal{E}_{ij}
  \frac{\partial B(\mathbf{p})}{\partial p_j}\bigg|_{\mathbf{p}=\mathbf{p}_0}.
\end{equation}
The uncertainty of observable $A(\mathbf{p})$ can be calculated as
\begin{equation}\label{eq:}
  \Delta A = \sqrt{C_{AA}}.
\end{equation}
The Pearson product-moment correlation coefficient between observables is evaluated as
\begin{equation}\label{eq:app-cor}
  c_{AB} \equiv \frac{C_{AB}}{\sqrt{C_{AA}C_{BB}}}.
\end{equation}

\begin{figure}
\includegraphics[width=8cm]{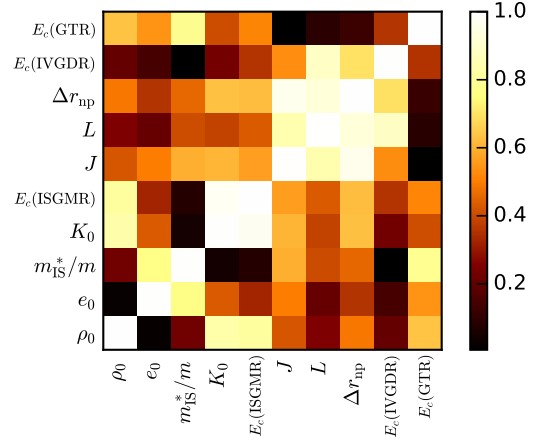}
\caption{(Color online) Pearson product-moment correlation coefficient matrix predicted by the covariance analysis of SAMi-T functional for various properties of nuclear matter and $^{208}$Pb (see text for the detail).}
\label{fig:app-cor}
\end{figure}

In Fig.~\ref{fig:app-cor} we show the Pearson product-moment correlation between different observables given by SAMi-T, including: saturation density $\rho_0$, saturation energy $e_0$, effective mass $m_{\rm IS}^*/m$, incompressibility coefficient $K_0$, centroid energy of GMR in $^{208}$Pb (Fig.~\ref{fig:gmr}), symmetry energy $J$, slope parameter of symmetry energy $L$, neutron skin $\Delta r_{\rm np}$ of $^{208}$Pb, centroid energy of IVGDR in $^{208}$Pb (Fig.~\ref{fig:gmr}), centroid energy of GTR in $^{208}$Pb (Fig.~\ref{fig:gt}).

Since the first five properties are of isoscalar nature and the second five are of isovector nature, there is a general pattern, as a rule, that the correlations between isoscalar and isovector properties are weaker than those among the same type.
For example, the correlations among saturation density, incompressibility, and centroid energy of the GMR are quite strong; the correlations among symmetry energy, slope parameter, neutron skin, and the centroid energy of the IVGDR are also strong.

In principle, the GTR excitation energy should be correlated with the parameter $G_0'$.
In the fit of SAMi-T, as well as in that of SAMi \cite{Roca-Maza2012}, this parameter is fixed to a reasonable value so that the GTR can be well reproduced.
Since $G_0'$ is fixed, this correlation does not exist for SAMi-T (or SAMi).
From Fig.~\ref{fig:app-cor} it can be seen that the GTR centroid energy is correlated with the effective mass.
This is understandable as the effective mass will influence the single-particle level density, which has an important effect on the GTR.

\section{Instability analysis}\label{app:stable}

\begin{figure}
\includegraphics[width=8cm]{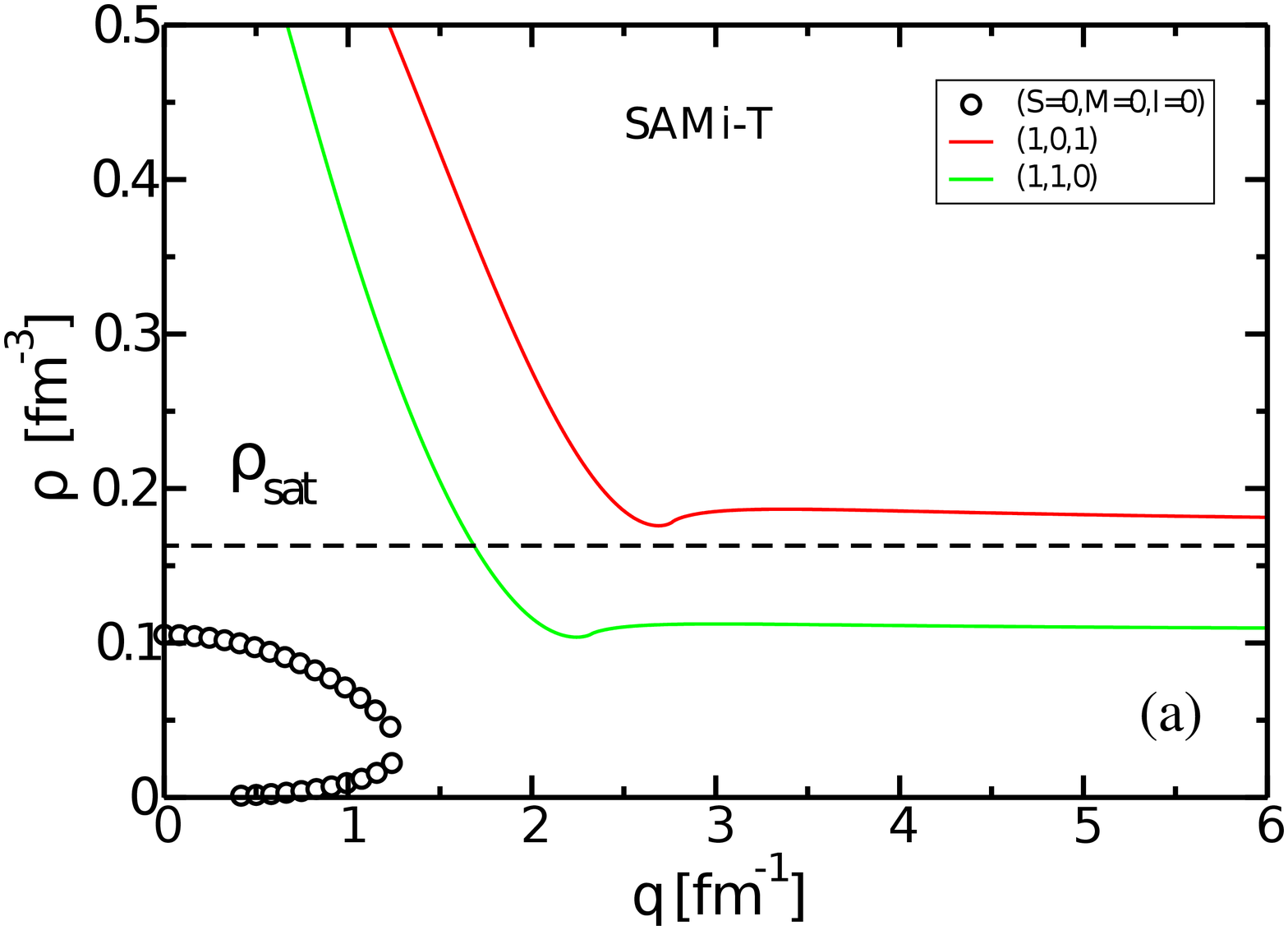}
\includegraphics[width=8cm]{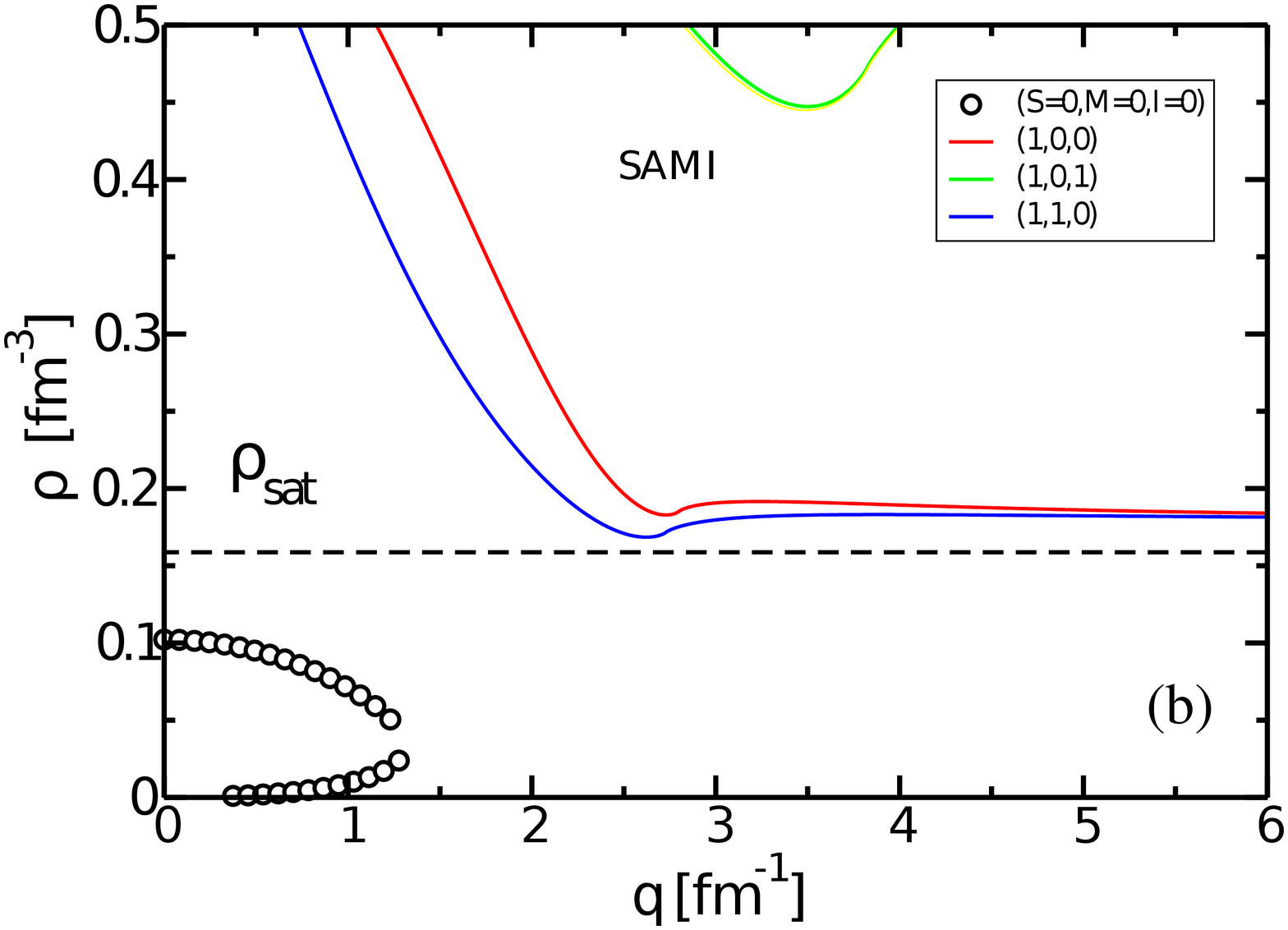}
\caption{(Color online) Position where instability occurs in the density $\rho$ and momentum transfer $q$ plane for (a) SAMi-T and (b) SAMi functionals.}
\label{fig:instability}
\end{figure}

It is known that the Skyrme functional exhibits spurious spin and spin-isospin instabilities, which can affect the properties of nuclear ground-state as well as excited-states \cite{Pastore2015}.
In Fig.~\ref{fig:instability}, we show the position where instability occurs in the density $\rho$ and momentum transfer $q$ plane for (a) SAMi-T and (b) SAMi functionals, obtained by solving the linear response function $\chi(q,\omega)$ in infinite nuclear matter.
The notations here are exactly the same as in Ref.~\cite{Pastore2015}.
Different channels are characterized by quantum numbers of the total spin $S$, projection $M$ along the $z$-axis, and total isospin $I$.
Beside the instability related to the gas-liquid phase transition in the $(S=0,M=0,I=0)$ channel, the others are unphysical.
It can be seen both SAMi-T and SAMi show instability near the saturation density, and the tensor term makes the functional more unstable.
But as the instability appears at a relatively large value of momentum transfer $q$, it shall not influence much on the discussions of this work.

%
\end{CJK}
\end{document}